# Fine structure of Surface Relief Gratings: experiment and a generic stochastic Monte Carlo model of the photoinduced mass transport in azo-polymer

G. Pawlik[a] and A.C. Mitus[b]
*Institute of Physics, Wroclaw University of Technology, Poland*
A. Miniewicz[c] and A. Sobolewska[d]
*Institute of Physical and Theoretical Chemistry, Wroclaw University of Technology, Poland*

[a]grzegorz.pawlik@pwr.wroc.pl, [b]antoni.mitus@pwr.wroc.pl,
[c]andrzej.miniewicz@pwr.wroc.pl, [d]anna.sobolewska@pwr.wroc.pl

We study experimentally and theoretically a double–peak fine structure of Surface Relief Gratings in azo–functionalized poly(etherimide). For the theoretical analysis we develop a stochastic Monte Carlo model for photoinduced mass transport in azobenzene functionalized polymer matrix. The long sought–after transport of polymer chains from bright to dark places of the illumination pattern is demonstrated and characterized, various scenarios for the intertwined processes of build–up of density and SRG gratings are examined. Model predicts that for some azo–functionalized materials double–peak SRG maxima can develop in the permanent, quasi–permanent or transient regimes.

Thin films of azobenzene functionalized polymers when exposed to interfering laser beams may develop a periodic surface corrugation pattern $\Delta d(x)$, called Surface Relief Grating [1,2] (SRG) below the polymer glass transition temperatures $T_g$. The microscopic polymer chain movements, which result from forces and torques due to light–induced multiple *trans–cis–trans* transitions of azobenzene moieties attached to the chains, induce a nonlinear macroscopic (a few micrometers) mass transport from bright to dark places of the illumination pattern at the surface of the polymer thin film. While studies of several different polymeric systems allowed for the progress in nanoscopic controlling of holographically inscribed surface patterns [3], the mechanisms responsible for the SRG formation are far from deep understanding [4]. Several theories have been formulated, including a mean-field model [5], pressure gradient model [6,7], competition between photoexpansion and photocontraction [8], viscoelastic flow model [9], stochastic inchworm–like motion [10], gradient force models [11,12], Navier–Stokes dynamics [13,14], random–walk model [15], stochastic models [16,17], light–induced softening [11,18], microscopic orientation approach [19,20,21,22] and others.

A rich variety of those approaches reflects various aspects of a fundamental mechanism of photoinduced mass transport. The physical picture is based on two processes. An inhomogeneous light illumination promotes the transport of polymer chains along the direction of the light intensity gradients (k–vector of the light interference grating) and a build–up process of the density grating along this direction (primary process). The emerging gradients of density, after exceeding a specific for the material threshold value, promote a surface corrugation and a build–up of SRG (secondary process). The dynamics of those intertwined processes is complex, the characteristic time scales, including time delay between build–up of both gratings, depend crucially on the polymer viscosity and elasticity, giving rise to various types of dynamics like, e.g., nearly simultaneous or strongly time–separated build–up of the density and SRG gratings.

A challenging issue arises: to develop a simple generic model which treats on an equal footing the polymeric matrix and functionalized dyes and reproduces main experimental observations. Such a "microscopic" model should be based on general arguments accounting for basic physical mechanisms related to dynamics/kinetics of the matrix and dyes. Important steps towards development of stochastic models of this kind were done in seminal papers of Nunzi *et al.*[10,4] and Juan *et al.*[16,17]. However these models, which successfully deal with the light–induced motion of photochromic dyes, do not account directly for the dynamics of polymeric chains and cannot study the directed mass transport of the polymer matrix which results in a build–up of SRG. The next step towards a generic model constitutes an extension of those theories and requires an inclusion of the polymer matrix at thermal equilibrium. A model of azopolymer host–guest systems, which directly accounts for the polymer matrix, was introduced and used in Ref. [23] for the Monte Carlo (MC) study of the kinetics of the inscription and erasure of diffraction gratings. A modification of this model, suitable for studies of functionalized dyes, was proposed in Ref. [24]. Preliminary studies have indicated that it is a promising candidate for a sought–after generic model of a photoinduced transport of functionalized polymer chains but no systematic studies were done.

The objective of the paper is twofold: (i), to fully develop a stochastic MC model for the kinetics of the photoinduced build–up of density/SRG gratings in a model polymer matrix functionalized with photochromic molecules and (ii), to use this model for a theoretical interpretation of latest experimental findings of the fine structure of SRG – an effect admittedly observed but either not discussed[25] or disregarded[26].

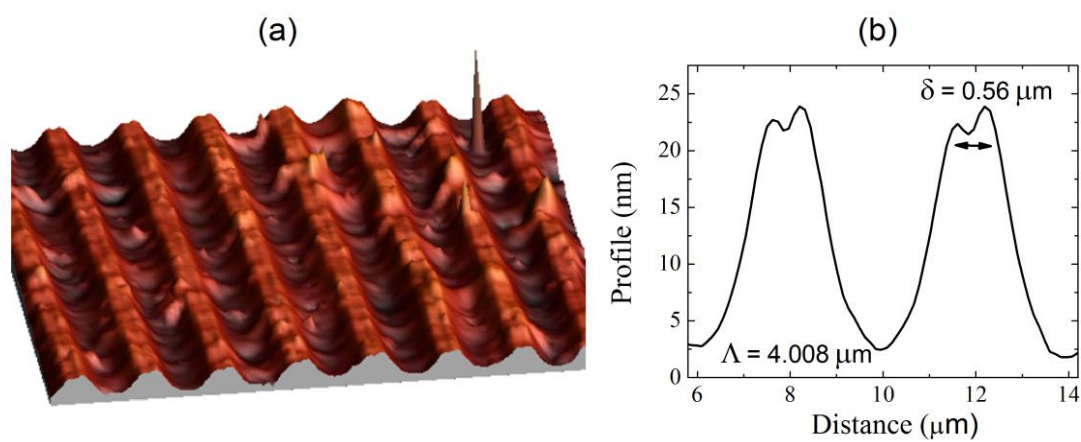

Fig 1. (a) 3-D plot of non-sinusoidal surface relief in azo–functionalized poly(etherimide) as measured by the AFM technique using a tapping mode. (b) Enlarged view of the SRG profile with the main periodicity $\Lambda = 4.0\,\mu$m and the peak-to-peak distance $\delta = 0.56\,\mu$m averaged over 22 scans performed along the grating wave vector. Grating was inscribed with $p-p$ configuration.

In Ref. [25] a non-sinusoidal pattern of SRG with a small valley at the top of the hills has been reported for the first time for a thin film of azo–functionalized poly(etherimide) in degenerate two-wave mixing (DTWM) experiment. The topography of the polymer surface $(30 \times 30)\mu m^2$ at the illuminated area was examined by atomic force microscopy (Dimension V scanning probe microscope, Veeco) working in a tapping mode. The 3-D view of the obtained structure is shown in Fig. 1 (a). We have re-analyzed the strongly fluctuating single irregular profiles of Ref. [25] and have performed a careful averaging of the corrugation structure profile for 22 new scans. The smooth profiles shown in Fig. 1 (b) have valleys of the width of $0.56\,\mu m$ at the top of the hills separated by $4\mu m$. Experimental details are as follows. The holographic DTWM grating recording was realized using a coherent linear polarization laser light ($\lambda = 514.5$ nm) from an argon gas laser (Innova 90, Coherent). The recording beam intensities were set equal amounting to $I_0 = 560$ mW/cm$^2$ each. The angle between writing beams was fixed at $\theta = 7.35^o$, resulting in a grating period of $\Lambda = 4.0\,\mu m$. Diffraction gratings were recorded for $p - p$ polarization configuration, in which the $\vec{E}$-vector of the light lies parallel to the light incidence plane, and $s - s$ one with the $\vec{E}$-vector lying perpendicularly to the incidence plane. The light intensity distribution along the grating period $I(x) = 2\,I_0\,(1 + \cos(2\pi x/\Lambda))$ for the two cases is sinusoidal and almost identical. The grating shown in Fig. 1 was inscribed using $p - p$ polarization for the exposure time of 30 min and the total first–order diffraction efficiency amounted to nearly $\eta = 1.3\%$. $s - s$ polarization configuration was used as well, reproducing a similar valley–like structure. The experiments for each polarization configuration were performed twice, giving reproducible diffraction efficiency dynamics and magnitude.

We conclude that the mechanism of the mass transport close to the surface is nonlinear and, at certain experimental conditions, a sub–wavelength resolution features can be generated. In Ref.[26] a clearly visible double peak structure is present as a transient effect, see corresponding Figs. 2(i) and 4(a); the grating was inscribed using s–s polarization geometry. This effect was left totally uncommented by the Authors. The absence of similar patterns in the literature probably results from the fact that no systematic observations of dynamics of the grating formation have been performed in a function of the exposure time so far. Additionally, suitable requirements for the polymer viscosity and elasticity must be met to observe the effect; for some systems it can become weak and/or transient, thus hard to observe or interpret[26].

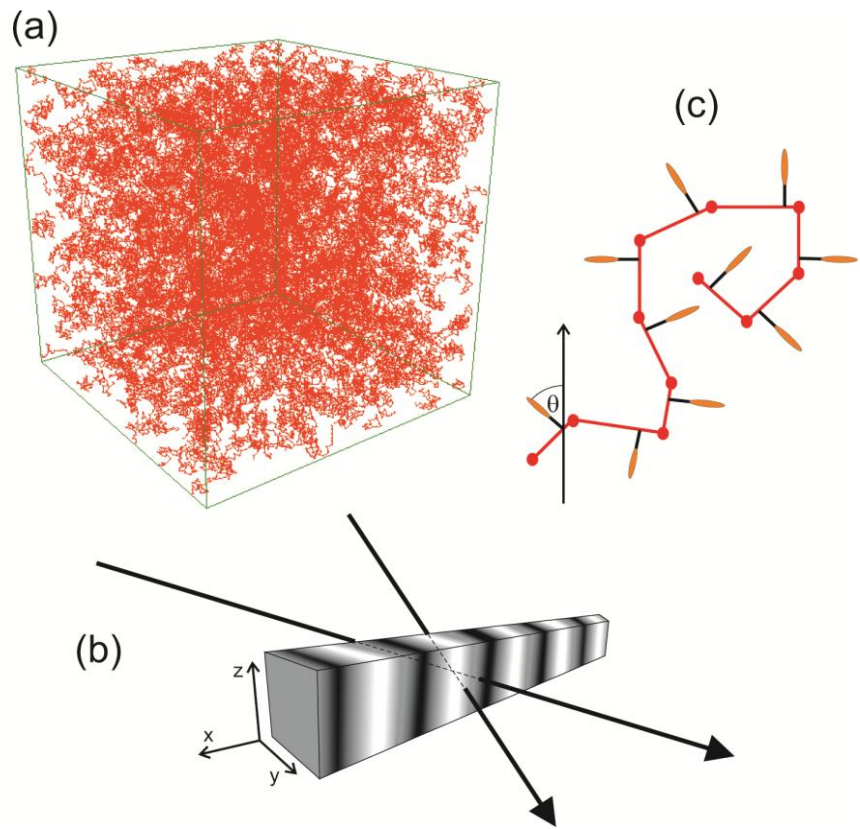

Fig. 2. Model of azo–functionalized polymer system. (a) Instantaneous configuration of 10% of all polymeric

chains; (b) illumination pattern $I(x)$ resulting from DTWM; (c) model of the bond–fluctuation polymeric chain with rigidly attached azo–dyes.

Below we give a short account of the stochastic Monte Carlo model. The polymer matrix was simulated using the bond–fluctuation MC method in 3D [27]. $N = 24000$ polymer chains, each consisting of $L = 20$ monomers, were placed on a $V_p = 200 \times 200 \times 200$ lattice forming a host system with monomer reduced density $\rho_0 = 8\,N\,L/V_p = 0.48$. Periodic boundary conditions were used for the study of the density grating and a free surface – for the SRG grating. The temperature was close to $T_g$. In a single Monte Carlo Step (MCS) each of the monomers performed a trial movement which was accepted when (i), a length of a trial bond did not violate imposed restrictions, (ii), steric constraints were obeyed and (iii), the Metropolis acceptance rule was met. More details can be found in Ref. [23]. The length of a typical run was $T = 10^5$ MCS. Fig. 2(a) shows the plot of an instantaneous configurations of polymeric chains.

We model the coupling of the optical field $I(x)$ resulting from DTWM (Fig. 2(b)) with the polymer matrix with attached chromophores as follows[24]. The dyes in $trans$ state are assumed to be strictly perpendicular to the bond (Kuhn element). The transition $trans \rightarrow cis$ occurs with the absorption probability proportional to $I(x)\cos^2\theta$, where $\theta$ stands for an angle between the long axis (transition dipole moment) of the $trans$ molecule and the direction of the light polarization. We assume, for simplicity, that the mechanical forces and torques exerted by the dyes on the chain during the photoizomerization reaction are independent on the angle $\theta$ ($\theta \neq \pi/2$) and are mimicked in MC modelling by an additional non-thermal trial movement of a monomer closest to the dye with the probability $0 \leq I(x)/2\,I_0 \leq 1$. The movement is accepted when the conditions (i) and (ii) (but not (iii)) are met. This way of the light–matter interaction preserves the left–right symmetry for the motion of a monomer along the direction of the modulation of the light field. The qualitative predictions of the model are polarization-independent ($s-s$, $p-p$) in agreement with experiment[25]; the quantitative differences are negligible.

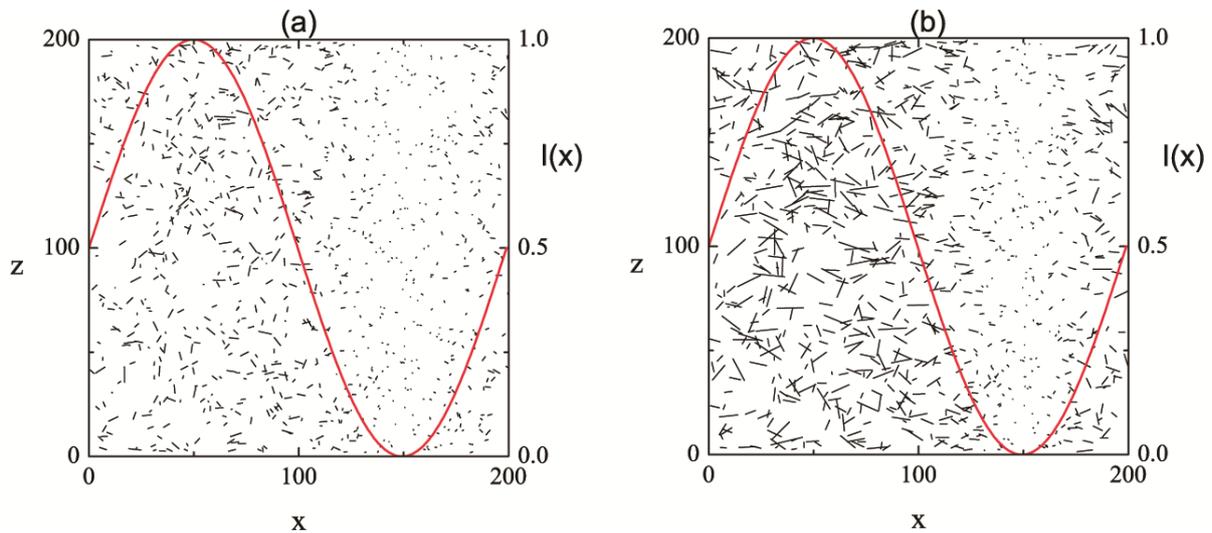

Fig. 3. (color online) Projections onto $x-z$ plane of the segments covered by the centers of mass of $10^3$ chains in the first $5 \times 10^3$ MCS (a) and $2 \times 10^4$ MCS (b), with superimposed illumination pattern $I(x)$ (red solid curve).

First, we study the primary process: build–up of the density grating. For constant illumination $I(x) = I$ the chains do not perform any directed motion. For the modulated intensity $I(x)$ an averaged pattern of the chain mobility shows a strong systematic dependence on the magnitude of the local illumination, see Fig. 3 which shows the projections onto $x - z$ plane of the segments covered by the centers of mass (CM) of chains in the initial phase of build-up of the density grating. In bright places, which correspond to the maximum of the illumination pattern, the large displacements are visible, while almost no movement is detected in dark places. The length of the segment is directly proportional to the absolute value of the average velocity of CM of a single chain in the whole simulation. The directed mass transport along $x$ axis in MC–time interval $(0, t)$ is characterized by the "macroscopic", i.e. averaged over chains, velocity of CM of a single chain:

$$v(x,t) = \frac{1}{N(x)\,t} \sum_i \left( x_i^{(CM)}(t) - x_i^{(CM)}(0) \right), \tag{1}$$

where $x_i^{(CM)}(t)$ denotes the $x$-component of the vector of CM of $i$-th chain at the time $t$. The summation is over those chains $i$ which at the time $t = 0$ had their $x$-component of CM at $x$: $x_i^{(CM)}(0) = x$; $N(x)$ denotes their number. Fig. 4 characterizes the average velocity $v(x, T/2)$ in the first half of the simulation, in the most active phase of build-up of the density grating. In bright places $v(x, T/2)$ is small: the absolute values of velocities are large (c.f. Fig. 3) but their directions are random. In dark places $v(x, T/2) \approx 0$. In the interval between the maximum of $I(x)$ at $x_{max} = 50$ and the minimum at $x_{min} = 150$ the velocity is positive which proves that there is a directed mass transport from bright to dark places. Moreover, $v(x, T/2)$ can be well fitted with the derivative of the illumination $I(x)$ (solid blue line). This observation holds also for other values of $t$. Thus, in our "microscopic" model

$$v(x,t) \propto -\frac{dI(x)}{dx}, \tag{2}$$

which is a phenomenological relation postulated on the macroscopic level[6,7,11,12]. Fig. 5 shows the monomer density profile $\rho(x)$ at the end of the simulation. Initially, the system was homogeneous: $\rho(x) = \rho_0 = 0.48$. One observes a decrease of the density in strongly illuminated places and its increase in weakly illuminated places. Comparison with properly scaled sine function (blue solid curve) shows that the profile $\rho(x)$ is close to sinusoidal with the exception of its high–density part which displays a weak, but well-defined minimum at $x_{min}$. Phenomenologically, this double–peak structure originates from two mass currents running in opposite directions: one due to the gradient of the illumination (Eq. (2)) and the second – to Fick's first law of the mass transport. Two small peaks close to $x_{min}$ build–up around places where both currents compensate each other. While this effect is a general rule, its magnitude depends on specific parameters of the polymer matrix as well as on the polarization setup[28] and can be negligibly small.

As discussed above, density grating promotes build–up of SRG. To model this effect we have simulated a system with a free surface which was stabilized by the surface–tension like force[24]. Its magnitude is an important parameter which is directly responsible for the time delay between build–up of the density and of SRG gratings. Fig. 6 shows emerging SRG profiles in two limiting cases. When stabilizing forces are weak both processes occur nearly simultaneously, the SRG profile (blue circles) is non–sinusoidal, without valleys at the top, in agreement with experimental studies of SRG in azo–functionalized poly(esterimide)[25]. The case of strong stabilizing forces is different: both processes are separated in time and SRG profile is built upon a well–developed density profile $\rho(x)$. In this regime SRG is nearly sinusoidal (thick red solid line), displays a double–peak structure and bears a strong resemblance to its parental density grating (Fig. 5) and the experimental SRG profile (Fig. 1 (b)). Double peak structure is a transient effect which smoothes away in later stages of the simulation. Nevertheless, in real experiments the corresponding time scales can be large and the double peak structure can be observed as a quasi–permanent or transient[26] effect. A detailed study of those processes will be published elsewhere.

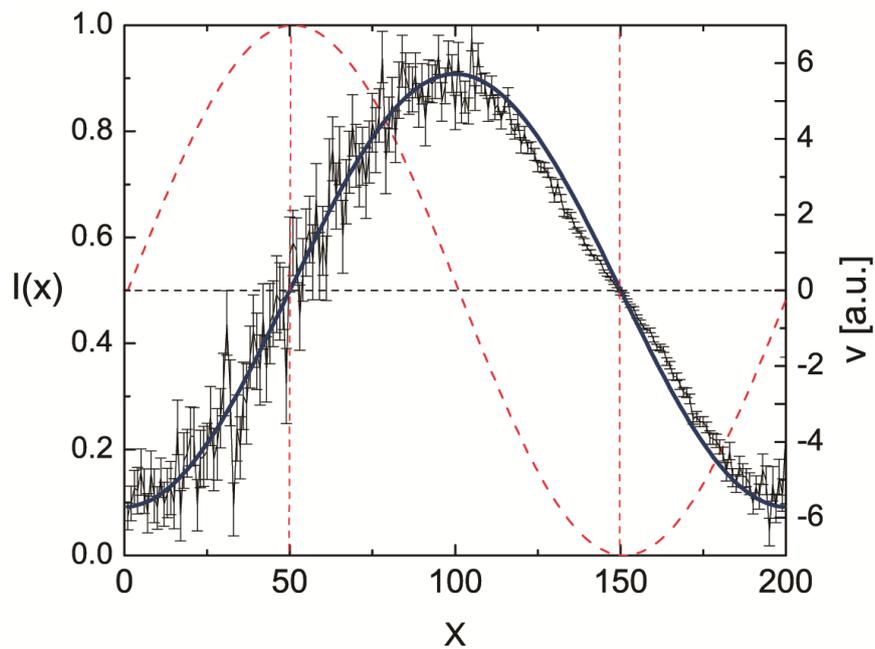

Fig. 4. Plot of the macroscopic velocity $v(x, T/2)$ of the center of mass of a chain, fitted with the (scaled) gradient of the illumination $I'(x)$ (blue solid curve); the illumination $I(x)$ (red dashed line).

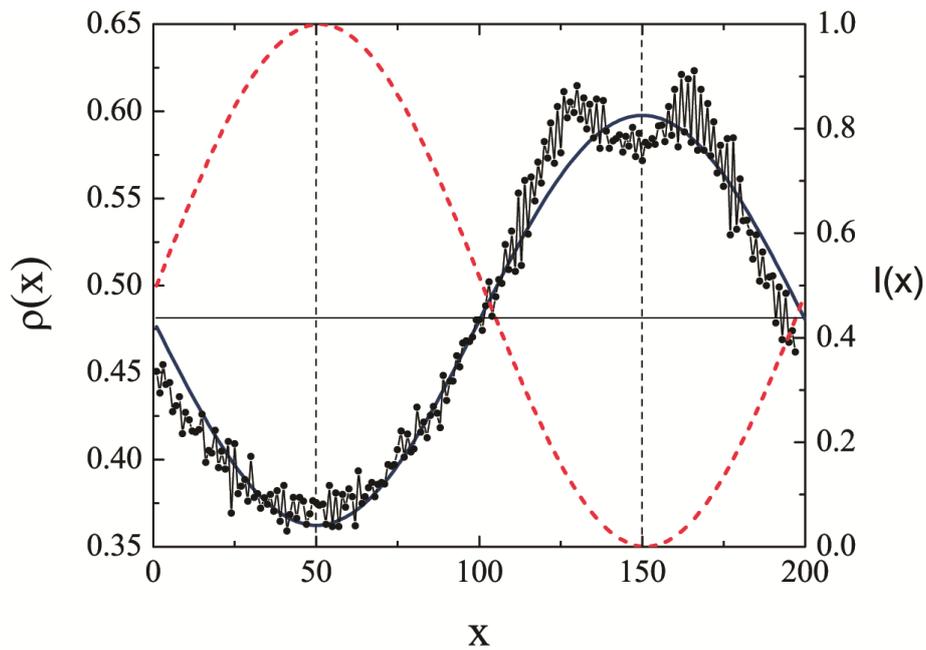

Fig. 5. Monomer density profile $\rho(x)$ at the end of the simulation and the plot (scaled) of a sine function (blue solid curve); the illumination $I(x)$ (red dashed line).

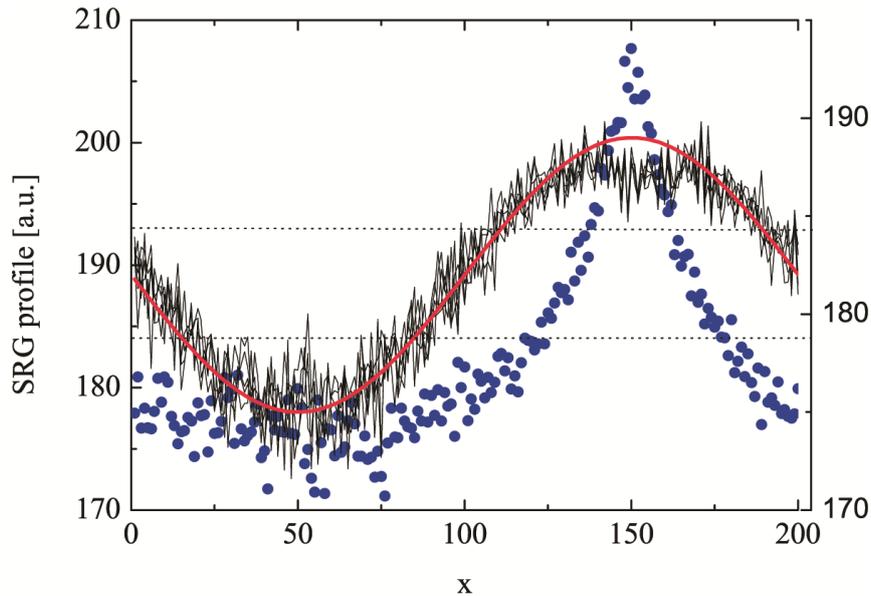

Fig. 6. SRG profile in a system with a free surface: weak stabilizing forces, end of the simulation (blue circles, scale left) and strong stabilizing forces, the initial phase of build–up of SRG (thin black solid line, scale right). Thick red solid line: scaled sine function, dashed horizontal lines: the surface before light illumination started.

The simple generic model of the mass-transport put forward in this Letter is weakly sensitive to the light polarization. However, it can be easily modified to account for polarization effects which substantially increases the area of its applications. For example, it can be applied to the theoretical study of the results of the elegant experiments on the recording of spiral–shaped reliefs in azo–polymer film using laser beam carrying the orbital angular momentum reported recently in Ref. [29]. The recording preserving handedness of the optical vortex has been explained by the anisotropic light–driven molecular diffusion[30]. As in our case, Authors of Ref. [30] conclude that the time correlation between orientational effects and mass transport effects could be responsible for the observed discrepancies between the theory and experiments.

To conclude, we have studied experimentally and theoretically the novel experimental effect – the build–up of the double–peaked SRG profiles observed in azobenzene functionalized polymers as permanent[25] or transient[26] effects. Its presence or absence in experimental studies is soundly explained by two scenarios put forward by the generic stochastic Monte Carlo model devised for the theoretical modelling of the photoinduced mass transport in a polymer matrix with functionalized azo-dyes.

**Acknowledgements**
A.M. and A.S. thank the Polish National Science Center for financial support of the project awarded on the basis of the Decision No. DEC-2011/03/B/ST5/01021. G.P. thanks the Polish National Science Center for financial support under Grant NN507 322440.